**Quantifying the effect of ionic screening with protein-decorated graphene transistors**

Jinglei Ping,[†] Jin Xi,[‡] Jeffery G. Saven,[§] Renyu Liu,[‡] & A. T. Charlie Johnson [*,†]

[†]Department of Physics and Astronomy, University of Pennsylvania, Philadelphia 19104, United States

[§]Department of Chemistry, University of Pennsylvania, Philadelphia, Pennsylvania 19104, United States

[‡]Department of Anesthesiology and Critical Care, University of Pennsylvania, Philadelphia, Pennsylvania 19104, United States

**ABSTRACT:** Liquid-based applications of biomolecule-decorated field-effect transistors (FETs) range from biosensors to in vivo implants. A critical scientific challenge is to develop a quantitative understanding of the gating effect of charged biomolecules in ionic solution and how this influences the readout of the FETs. To address this issue, we fabricated protein-decorated graphene FETs and measured their electrical properties, specifically the shift in Dirac voltage, in solutions of varying ionic strength. We found excellent quantitative agreement with a model that accounts for both the graphene polarization charge and ionic screening of ions adsorbed on the graphene as well as charged amino acids associated with the immobilized protein. The technique and analysis presented here directly couple the charging status of bound biomolecules to readout of liquid-phase FETs fabricated with graphene or other two-dimensional materials.

Introduction

Graphene, a single-atom thick layer of sp$^2$ carbon, is a promising solid-state material for use as an interface to biological systems. Graphene has outstanding biocompatibility[1,2], and may be fabricated into graphene field effect transistors (GFETs) suitable for decoration with biomolecules such as proteins, or even host active neural cells[1]. In aqueous solution, biomolecules decorating the GFET channel may acquire charges, which will in turn affect the carrier density that governs electron transport through the graphene. When the system is measured in liquid, the electric field associated with charge on the biomolecules is screened by mobile ions and polarization of water molecules; a quantitative model for this effect is essential for understanding the biomolecule/graphene system[3,4]. Prior investigations of this effect were based on either nanowire FETs[3], where the channel geometry is not well defined and device yield is relatively low[5], or single-molecule carbon nanotube FETs[6-8] that are not well-suited for scalable fabrication. These elegant studies were mainly performed by measuring real-time fluctuations of the current (or equivalently, resistance or conductance) through the channel due to conformational motion of a single bound protein. The ionic screening effect was understood by the Debye-Hückel model but the effect of protein charging on FET transport properties was treated as an experimentally determined parameter.

Here we present a detailed analysis of current-back gate voltage ($I$-$V_{bg}$) measurements of bare and protein-decorated GFETs in phosphate buffered saline (PBS). We tuned the Debye screening by varying the PBS ionic strength $c$ from 0.3 – 150 mM, where the high end of this range is similar to that of complex biofluids, such as cerebrospinal fluid or blood. The pH was precisely controlled at 7.0 because the conductivity of graphene is known to be sensitive to this variable[9,10]. Observed changes in the $I$-$V_{bg}$ relationship of bare graphene devices with respect to ionic strength agree well with existing theories of the electronic double layer that exists at the graphene-ionic solution interface. For protein-functionalized GFETs, we develop a *quantitative* model linking the transport properties of the graphene with the charge that accumulates on proteins bonded to the GFET surface and find excellent agreement with the data. We find that the electric field of biomolecular charges chemically gates[11] the graphene, altering its transport properties. Thus we confirm that measured changes in the GFET $I$-$V_{bg}$ are due to electrostatic chemical gating effect[11] rather than Faradaic charge transfer[12,13]. The techniques and analytical methods developed here represent significant progress towards quantitative understanding of nano-enabled sensor systems and may be generalized for liquid-phase measurements of protein-functionalized biosensors based on other two-dimensional materials such as $MoS_2$ and $WS_2$[14,15].

Material and Methods

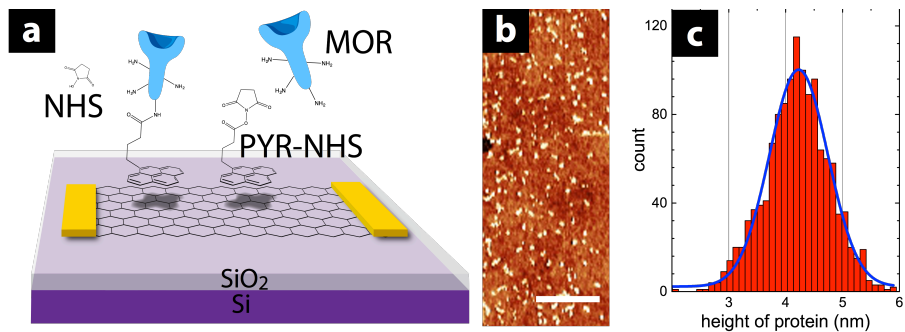

Figure 1. (a) Schematic of chemistry used to immobilize water-soluble mu-opioid receptors (wsMORs) on graphene with 1-Pyrenebutyric acid N-hydroxysuccinimide ester ("PYR-NHS"). (b) Atomic force micrograph shows the high efficiency of the immobilization. The scale bar is 0.5 µm. (c) Histogram of measured heights of wsMORs obtained from the AFM images along with a Gaussian fit (blue line).

We use a low-contamination fabrication method[16] that minimizes contact between the transferred graphene layer and chemicals typically used in device patterning, to avoid negative impact of lithographic processing on electrical and chemical properties of graphene as well as its biological compatibility (See Supporting information for details of Methods). The GFETs have very good transport properties with mobility ~ 2000 cm$^2$ V$^{-1}$ s$^{-1}$. We functionalized the GFETs with a computationally redesigned, water-soluble variant of the human mu-opioid receptor (MOR) variant, referred to as wsMOR[17,18]. The wsMOR has a 3D-structure similar to that of the wild-type receptor as well as comparable opioid affinities, even at ionic strength as low as 0.3 mM[16]. Observed changes in graphene transport properties are attributed to the variation of electrostatics caused by wsMOR binding. We use 1-Pyrenebutyric acid N-hydroxysuccinimide ester (PYR-NHS, Sigma-Aldrich) as a bifunctional linker[19,20] to functionalize GFETs with wsMOR (See Methods in Supporting information). Atomic Force Microscopy shows that the density of wsMOR bound to the GFET surface is ~ 115 μm$^{-2}$ (Fig. 1b). The height histogram of bound wsMOR (Fig. 1c) is well fit by a Gaussian distribution with mean value ~ 4.2 nm, consistent with the known molecular weight (~ 46 kDa) and structure of wsMOR.

Results and Discussion

In the framework of the electrical double layer (EDL)[21], when bare graphene is in ionic solution, a plane of negative ions[9,22], the Stern plane, adsorbs such that the nuclei are separated from the graphene by a "Stern layer" with thickness $\xi$ ~ 0.3 – 0.4 nm[9,10,23], where the molecular density of water is lowered by 97% compared to the bulk[24-26]. The Stern plane acts as a chemical gate that induces a compensating charge density in the graphene and causes a shift in the Dirac voltage, $V_D$, the back-gate voltage where the conductance of the graphene is a minimum. Beyond the Stern layer, there is a diffuse layer of ions where the electric and osmotic potentials are balanced. The electric potential $\psi$ decays exponentially with distance into the solution in the diffuse layer, $\psi = \psi_d e^{-(z-\xi)/\lambda_D}$, where $z$ is the distance from a point in the solution to the graphene, $\psi_d$ is the potential at the Stern plane, and $\lambda_D$ is the Debye screening length (See Supporting information for details of the calculation of $\lambda_D$).

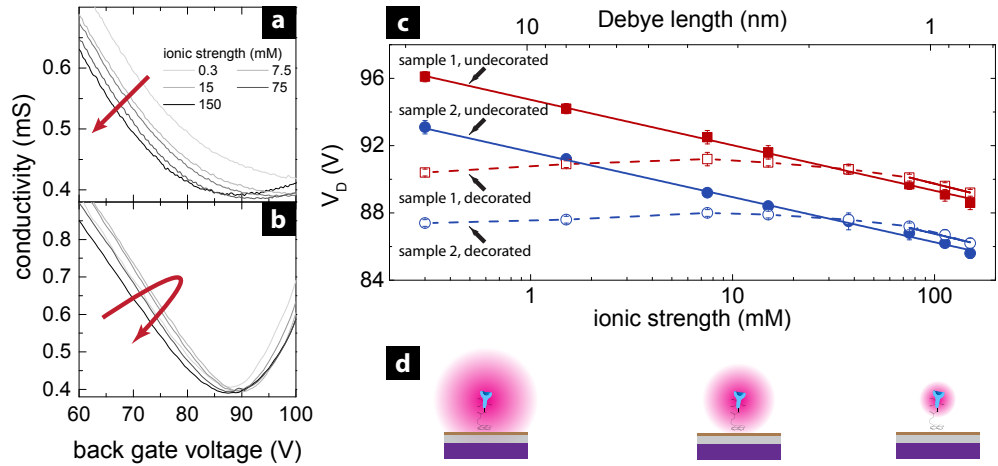

Figure 2. (a) Conductivity vs. back gate voltage for undecorated GFET sample 1. The Dirac voltage decreases monotonically with ionic strength $c$. (b) Conductivity vs. back gate voltage for the same device decorated with wsMOR. The Dirac voltage behavior is non-monotonic. Arrows in (a) and (b) indicate the shift of Dirac voltage as $c$ is increased. (c) Dirac voltage vs. ionic strength for two devices (Sample 1 is featured in panels (a) -(b)). Solid lines are fits to the data with formalism described in the Section 4 of the Supporting Information. (d) Schematic illustration of ionic screening of the electrostatic field due to charges on wsMOR, at ionic strengths (Debye screening lengths) indicated on the bottom (top) x-axis of panel (c).

We explored ionic screening effects for bare and wsMOR-decorated GFET devices through measurements of the Dirac voltage in solution with varying ionic strength (See Supporting information for Dirac voltage extraction). For undecorated GFETs, the inferred value of $V_D$ (solid points in Fig. 2a,b,c) decreases with ionic strength with a characteristic log-linear behavior that reflects increased ionic screening of the charges in the Stern layer[9,10,23,25] (See Section 1 of the Supporting Information for additional details). Strikingly, for wsMOR-decorated GFETs the Dirac voltage exhibits a different, *non-monotonic* dependence on ionic strength (Fig. 2b,c). The Dirac voltage increases with ionic strength until a critical "knee" point at ionic strength ~ 10 mM ($\lambda_D$ ~ 3 nm), after which the slope of the curve decreases and changes sign from +0.6 to -3.0 (Fig. 2c). Undecorated or decorated, samples 1 and 2 show identical behavior except for an offset in $V_D$ that is ascribed to different charge trap densities for the two substrates.

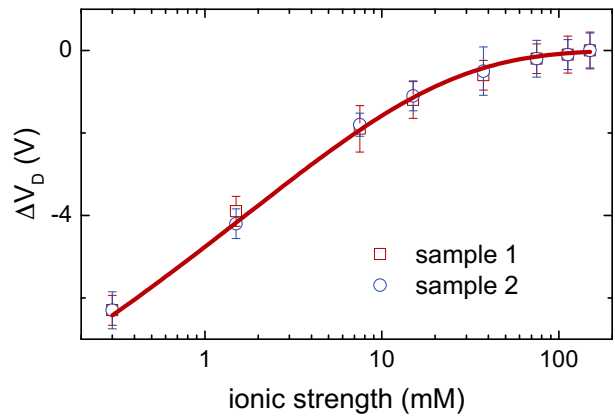

Figure 3. Difference between the Dirac voltages for two devices, before and after decoration with wsMOR. The saturation value of $\Delta V_D$ (see Fig. 2c) has been suppressed. The red solid curve is a fit to the data using Eqn. (1) from the main text.

Proposed mechanisms for these observations involve protonated residues on the wsMOR (~ 70 total), which provide either (Faradaic) electron transfer[13] or chemical gating due to protonation[27]. At low ionic strength (0.3 mM), the difference in Dirac voltage between undecorated and decorated versions of the same device, $\Delta V_D$ ~ -6 V (Fig. 2c and 3), corresponding to carrier density of 4000 e⁻ µm⁻², or ~ 35 electrons per bound wsMOR. The Faradaic model requires a charge transfer efficiency of ~ 55%, which is implausibly large and greater by an order of magnitude than reports of 4% for primary amine groups near carbon nanotube devices[12,38]. In contrast, this carrier density is readily explained quantitatively by considering the combined effect of basic and acidic protein residues subject to protonation and deprotonation, respectively, in solution. To estimate the net charge of the wsMOR, we consider pKa values for the relevant amino acids in the free state (44 lysine, 15 arginine, 14 histidine, 8 aspartic acid , and 25 glutamic acid ) and find a net of ~ + 38 fundamental charges per wsMOR, in good agreement with the data.

To our knowledge the problem of chemical-gating effect by charged biomolecules in ionic solution has not been solved analytically. We therefore analyze the problem approximately by assuming that the graphene acts as a perfect electrical conductor, and the protein molecule has a spherically symmetric charge distribution whose electric field decreases exponentially into the diffuse layer due to ionic screening on the scale of the Debye length. In this approximation, the Dirac voltage change $\Delta V_D$ induced by wsMOR decoration can be found with the image charge technique[28,29]:

$$\Delta V_D = -\frac{4\pi\varepsilon\varepsilon_0 \sigma_A}{C_{ox}} \frac{2k_B T}{e} \sinh^{-1} \frac{\rho_M}{\sqrt{8\varepsilon\varepsilon_0 k_B T c}} a e^{-\frac{a}{\lambda_D}} \quad , \quad (1)$$

where $\sigma_A$ is the wsMOR density of 115 μm$^{-2}$, $\rho_M$ is the net surface charge density of the wsMOR, and $a$ is the distance from the graphene to the center of the wsMOR molecule. This result is analogous to the Debye-Hückel model for screening of a point charge by ionic solution[6,30]. We note that $\Delta V_D$ depends exponentially on the Debye length, in agreement with the observation that the value of $\Delta V_D$ tends towards zero for ionic strength ~ 10 mM, where $\lambda_D \approx a$ and the knee point of $V_D$ for the decorated GFETs as shown in Fig. 2c.

As shown in Fig. 3b, the data for $\Delta V_D$ as a function of ionic strength $I$ are very well described by Eqn. (1), where the fit parameters are the distance $a$ and the charge density $\rho_M$; the best fit parameter values for $a$ and $\rho_M$ are 3.5±0.5 nm and 0.11±0.02 C m$^{-2}$ respectively. Based on the coupling chemistry used in the experiment, $a$ is expected to be ~ 3.1 nm, in good agreement with the fit. With the approximation that the radius of the ion-shell outside the wsMOR is approximately the radius of wsMOR plus the thickness of the Stern layer (a total of ~ 2.3 nm), the best fit value of the charge density $\rho_M$ corresponds to 46 ± 9 positive charges per wsMOR molecule, in good agreement with ~ 38 positive sites per wsMOR that was estimated above by considering charging of the wsMOR residues in solution.

Conclusions

We used fabricated clean and bio-compatible GFETs and decorated them with wsMOR using a bifunctional linker molecule. We measured the Dirac voltage shift as a function of ionic strength and provided a quantitative explanation based on accepted theory for the electric double layer and a simplified model of protein electrostatics. The results are fully consistent with the action of solution-mediated chemical-gating by bound wsMOR molecules as the mechanism that shifts the Dirac voltage of graphene in our biosensor system. This investigation shows the potential of this system for studying the charging properties of biomolecules bound to the FET channel based on graphene or other two-dimensional materials.

**AUTHOR INFORMATION**

**Corresponding Author**

cjohnson@sas.upenn.edu

**Funding Sources**

This work was supported by the Defense Advanced Research Projects Agency (DARPA) and the U. S. Army Research Office under grant number W911NF1010093. Additional support is acknowledged from: National Science Foundation Accelerating Innovation in Research program AIR ENG1312202, the Nano/Bio Interface Center NSF NSEC DMR0832802, FAER (Foundation for Anesthesia Education and Research, PI, RL), NIH K08 (K08-GM-093115-01) (PI, RL), NIH R01 1RO1GM111421-01, and GROFF (PI, RL), and the Department of Anesthesiology and Critical Care at the University of Pennsylvania (PI, RL).

**ACKNOWLEDGMENT**

JGS acknowledges infrastructural support from the Penn LRSM MRSEC (NSF DMR1120901).